\documentclass[sigconf,10pt,review=false,nonacm]{acmart}
\usepackage{tikz}
\usetikzlibrary{arrows.meta}
\usetikzlibrary{automata,positioning}
\usetikzlibrary{calc}
\usetikzlibrary{arrows.meta,arrows,backgrounds,patterns,positioning,shadings, decorations.markings, shapes.multipart, shapes}
\usepackage{color}
\usepackage{graphicx,subcaption,ragged2e}
\usepackage{soul}

\title{LLM-based Config Synthesis requires Disambiguation}

\author{Rajdeep Mondal}
\affiliation{%
  \institution{University of California Los Angeles}%
  \city{Los Angeles, CA}
  \country{USA}%
}

\author{Nikolaj Bjorner}
\affiliation{%
  \institution{Microsoft Research}%
  \city{Redmond, WA}
  \country{USA}%
}

\author{Todd Millstein}
\affiliation{%
  \institution{University of California Los Angeles}%
  \city{Los Angeles, CA}
  \country{USA}%
}

\author{Alan Tang}
\affiliation{%
  \institution{Microsoft}%
  \city{Redmond, WA}
  \country{USA}%
}

\author{George Varghese}
\affiliation{%
  \institution{University of California Los Angeles}%
  \city{Los Angeles, CA}
  \country{USA}%
}

\newcommand{\matches}{\mathit{matches}}
\usepackage{listings}
\definecolor{backcolour}{rgb}{0.95,0.95,0.92}
\lstdefinelanguage{cisco}{
    alsoletter={-},
    morekeywords = [1]{match,set,permit,deny,continue,goto},
    morekeywords = [2]{route-map,ip,prefix-list,community-list,access-list},
    morekeywords = [3]{extended,additive,seq,address,community},
    keywordstyle = [1]\color{orange},
    keywordstyle = [2]\color{magenta},
    keywordstyle = [3]\color{brown},
    sensitive = false,
    morecomment = [l]{!},
    morecomment = [s]{/*}{*/},
    morecomment = [s]{/**}{*/},
    commentstyle = \color{red}\bf,
    morestring = [b]",
    morestring = [b]',
    stringstyle = \color{purple}
}

\lstdefinestyle{mystyle}{
    language={cisco},
    basicstyle=\ttfamily, 
    breakatwhitespace=false,         
    breaklines=true,                 
    captionpos=b,                    
    keepspaces=true,                       numbers=left,
    numbersep=5pt,                  
    showspaces=false,                
    showstringspaces=false,
    showtabs=false,                  
    tabsize=2, 
    numbers=none
}

\lstdefinestyle{mystyle2}{
    language={cisco},
        basicstyle=\selectfont\ttfamily,
    breakatwhitespace=false,         
    breaklines=true,                 
    captionpos=b,                    
    keepspaces=true,                       
    numbersep=5pt,                  
    showspaces=false,                
    showstringspaces=false,
    showtabs=false,                  
    tabsize=2,
    numbers=none
}

\lstdefinelanguage{json}{
    basicstyle=\normalfont\ttfamily,
    numbers=left,
    numberstyle=\scriptsize,
    stepnumber=1,
    numbersep=8pt,
    showstringspaces=false,
    breaklines=true,
    frame=lines,
    backgroundcolor=\color{background},
    literate=
     *{0}{{{\color{numb}0}}}{1}
      {1}{{{\color{numb}1}}}{1}
      {2}{{{\color{numb}2}}}{1}
      {3}{{{\color{numb}3}}}{1}
      {4}{{{\color{numb}4}}}{1}
      {5}{{{\color{numb}5}}}{1}
      {6}{{{\color{numb}6}}}{1}
      {7}{{{\color{numb}7}}}{1}
      {8}{{{\color{numb}8}}}{1}
      {9}{{{\color{numb}9}}}{1}
      {:}{{{\color{punct}{:}}}}{1}
      {,}{{{\color{punct}{,}}}}{1}
      {\{}{{{\color{delim}{\{}}}}{1}
      {\}}{{{\color{delim}{\}}}}}{1}
      {[}{{{\color{delim}{[}}}}{1}
      {]}{{{\color{delim}{]}}}}{1},
}

\newcommand{\tool}{{\sf Clarify}}
\newcommand{\remove}[1]{}

\newcommand*\mycirc[1]{%
  \begin{tikzpicture}[baseline=(C.base)]
    \node[draw,circle,inner sep=1pt, fill=blue!10](C) {#1};
  \end{tikzpicture}}

\begin{abstract}
Beyond hallucinations, another problem in program synthesis using LLMs is ambiguity in user intent.  We illustrate the ambiguity problem in a networking context for LLM-based incremental configuration synthesis of route maps and ACLs. These structures frequently overlap in header space, making the relative priority of actions impossible for the LLM to infer without user interaction.
Measurements in a large cloud identify complex ACLs with 100’s of overlaps, showing ambiguity is a real problem. We propose a prototype system, \tool, which uses an LLM augmented with a new module called a {\em Disambiguator} that helps elicit user intent. On a small synthetic workload, \tool{} incrementally synthesizes routing policies after disambiguation and then verifies them.
Our treatment of ambiguities is useful more generally when the  
intent of updates can be correctly synthesized by LLMs, but their integration 
is ambiguous and can lead to different global behaviors. 
\end{abstract}

\begin{document}
\maketitle

\section{Introduction}
\label{intro}
\textit{"Life is ambiguous; there are many right answers - all depending on what you are looking for."} - Roger van Oech
\vspace{0.1in}

While LLM technology for program synthesis is dramatically improving and hallucinations may disappear, LLMs will never be able to read a user's mind for their intent.
Techniques like RAG \cite{rag, graphrag, selfrag, agenticrag, speculativerag, modularrag}, chain of thought \cite{chain_of_thought} and Agentic AI~\cite{AgenticAI} greatly reduce 
hallucinations and incorrect output. However, one key bottleneck remains even if an LLM can perform program synthesis perfectly: the need for the user to fully and unambiguously specify their intent. This is difficult to do even for relatively simple settings and is infeasible to expect users to do correctly for realistic tasks.

For instance, a recent study ~\cite{berkeley} on disentangling possible meanings from ambiguous English sentences found that only 32\% of the LLM-proposed resolutions were considered correct in crowd-sourced evaluations. Another study~\cite{selfconsistency} showed that LLMs are inconsistent in applying factual knowledge when prompted with ambiguous entities, with performance deteriorating to 75\% with under-specified commands.

We present an approach that addresses this problem in the context of synthesis for program updates, where an existing piece of code is extended to support new functionality or fix bugs. For concreteness, we focus on updates to network configurations, specifically updates to routing policy (route-maps) and access control (ACLs). Such updates happen frequently and need to be correct, and while LLMs are a natural fit for synthesizing updates, the lack of unambiguous specifications remains a limiting factor in practice.

We observe that often the intent of an update is itself relatively simple and unambiguous. However, if care is not taken then this update can easily cause regressions and unexpected behavior through interactions and interference with existing parts of the configuration, based on where the update is inserted. The basic idea of our approach is to leverage this observation by asking the LLM to synthesize a config snippet \textit{in isolation} given the intent of a change, and then to use a new component that we call a \textit{disambiguator} to determine where to place the snippet in order to satisfy the full user intent. 

\begin{figure*}
\centering
\begin{tikzpicture}
    \node[inner sep=3pt,fill=white!7] (background) at (0,0) {
        \begin{tikzpicture}
            \node[](wf) at (0,0) {\includegraphics[width=\textwidth]{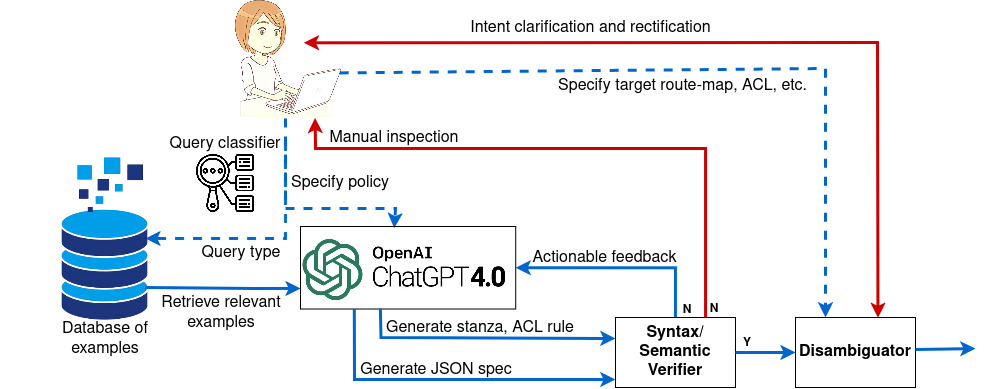}};
            \node[label={}] (circ1) at (-5.6, -1.1) {\mycirc{1}};
            \node[label={}] (circ2) at (-4, -2.5) {\mycirc{2}};
            \node[label={}] (circ2a) at (-1.2,-0.25) {\mycirc{2}};
            \node[label={}] (circ3) at (1,-3) {\mycirc{3}};
            \node[label={}] (circ4) at (2,-1.75) {\mycirc{4}};
            \node[label={}] (circ5) at (-0.2,1.2) {\mycirc{5}};
            \node[label={}] (circ6) at (6.5,1.8) {\mycirc{6}};
        \end{tikzpicture}};
\end{tikzpicture}
\vspace{-1em}
\caption{Incremental Synthesis with Verification: The user specifies an atomic policy that is fed into a query classifier to identify the query type \protect\mycirc{1}. Then the initial prompt along with the retrieved context are jointly supplied to the LLM \protect\mycirc{2}. It produces the relevant config snippet and a formal JSON specification \protect\mycirc{3}. The verifier checks the equivalence between the snippet and the formal spec and provides actionable feedback \protect\mycirc{4} to the LLM in case of errors. If automatic rectification fails, then the user steps in \protect\mycirc{5}. Afterwards, the snippet is inserted into the original configuration using disambiguation to resolve potential conflicts \protect\mycirc{6}.}
\label{workflow}
\end{figure*}
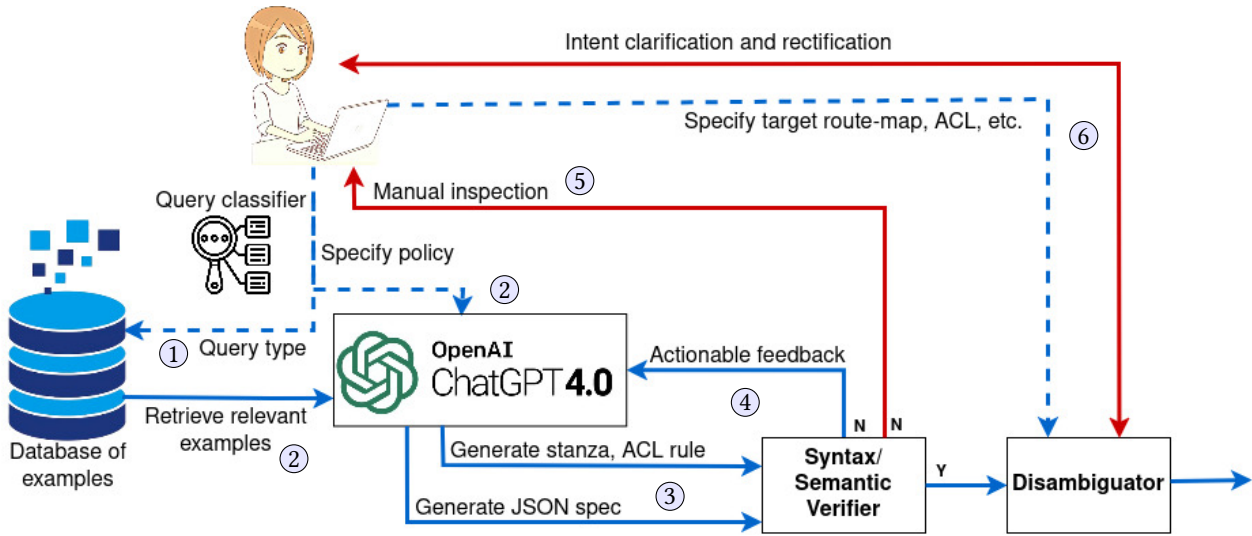

Figure~\ref{workflow} shows the overall flow diagram of our proposed approach. Like prior work on LLM-based network configuration synthesis~\cite{hotnets}, we iterate synthesis with verification. However, our approach performs synthesis {\em incrementally}: each synthesis call produces a single new stanza to add to an existing configuration policy (route-map or ACL). This stanza is specified and synthesized {\em in isolation}, which dramatically simplifies the jobs of both the user (in specifying behavior) and the LLM (in synthesizing a correct stanza).  We then introduce a new component called a {\em disambiguator}, which asks the user targeted behavioral questions to determine where to place the new stanza within the existing configuration, effectively eliciting the full specification from the user in an incremental fashion. 

\section{Disambiguation workflow}
\label{methodology}
In the cyclic workflow shown in Figure~\ref{workflow}, 
the user starts with a simple natural language intent to edit a route-map or an access-list without prior knowledge of possible conflicts. The intended behavior is then clarified through differential examples shown to the user over multiple interactions.

\subsection{LLM Query and Verification}
\label{query}
Consider the following route-map written in Cisco IOS:
\vspace{10pt}
{\color{black}\hrule}
\vspace{-4pt}
\begin{lstlisting}[style=mystyle]
ip as-path access-list D0 permit _32$

ip prefix-list D1 seq 10 permit 10.0.0.0/8 le 24
ip prefix-list D1 seq 20 permit 20.0.0.0/16 le 32
ip prefix-list D1 seq 30 permit 1.0.0.0/20 ge 24

route-map ISP_OUT deny 10
 match as-path D0
route-map ISP_OUT deny 20
 match ip address prefix-list D1
route-map  ISP_OUT permit 30
 match local-preference 300
\end{lstlisting}
\vspace{-4pt}
{\color{black}\hrule}
\vspace{10pt}
We want to add a new stanza to the above that permits routes with the community 300:3 and prefix 100.0.0.0/16 having mask lengths less than or equal to 23 and sets their metric to 55. So we provide the following prompt to the LLLM in simple English language (note that we refrain from including any \texttt{ISP\_OUT} specific information in the prompt):
\vspace{10pt}
{\color{black}\hrule}
\vspace{-4pt}
\begin{lstlisting}[basicstyle=\ttfamily\bfseries, breaklines=true]
Write a route-map stanza that permits routes containing the prefix 100.0.0.0/16 and tagged with the community 300:3. Their MED value should be set to 55.
\end{lstlisting}
\vspace{-4pt}
{\color{black}\hrule}
\vspace{10pt}


To prevent LLM errors, we augment the system prompt with
a task description (e.g., generate only one route-map stanza in Cisco IOS syntax) 
and few-shot examples containing similar prompts and their translations.

When the user issues the original query, we make an intermediate LLM call (\textit{the query classification step} \mycirc{1} \textit{in} Figure \ref{workflow}) to identify its type (whether it is an ACL synthesis query or a route-map synthesis query), and retrieve the appropriate system prompts and examples from a database (\mycirc{2} in Figure \ref{workflow}). This step also selects the 
synthesis pipeline, 
in this case a route-map synthesis query.

In response, GPT-4 produces (\mycirc{3} in Figure \ref{workflow}):
\vspace{10pt}
{\color{black}\hrule}
\vspace{-4pt}
\begin{lstlisting}[style=mystyle]
ip community-list expanded COM_LIST permit _300:3_
ip prefix-list PREFIX_100 permit 100.0.0.0/16 le 23

route-map SET_METRIC permit 10
 match community COM_LIST
 match ip address prefix-list PREFIX_100
 set metric 55
\end{lstlisting}
\vspace{-4pt}
{\color{black}\hrule}
\vspace{10pt}



To verify the generated snippet's correctness, 
we use GPT-4 again to 
to produce a JSON specification from the modified user prompt (\mycirc{3} in Figure \ref{workflow}) as:
\vspace{10pt}
{\color{black}\hrule}
\vspace{-4pt}
\begin{lstlisting}
<@\textbf{\textcolor{black}{\{}}@>
    "permit": true,
    "prefix": <@\textbf{\textcolor{black}{[}}@>"100.0.0.0/16:16-23"<@\textbf{\textcolor{black}{]}}@>,
    "community": "/_300:3_/",
    "set": <@\textbf{\textcolor{black}{\{}}@>
    	"metric": 55
    <@\textbf{\textcolor{black}{\}}}@>
<@\textbf{\textcolor{black}{\}}}@>
\end{lstlisting}
\vspace{-4pt}
{\color{black}\hrule}
\vspace{10pt}

The user checks the specification manually to ensure that it has semantics equivalent to the original intent, which for one stanza is easy to cross-check.
We then verify that the synthesized stanza meets the synthesized specification using existing Batfish analysis methods for proving behavioral properties of ACLs and route maps ({\tt searchFilters} and {\tt searchRoutePolicies}). 
The automated verification and feedback cycle continues until the LLM finally produces the correct output or we reach a threshold and punt to the user (\mycirc{5} in Figure~\ref{workflow}) who starts over or provides more information.

\begin{figure*}
\captionsetup[subfigure]{justification=Centering}

\begin{subfigure}[t]{0.45\textwidth}
{\color{black}\hrule}
\vspace{-4pt}
    \begin{lstlisting}[style=mystyle2]
ip community-list expanded D2 permit _300:3_
ip prefix-list D3 permit 100.0.0.0/16 le 23  
\end{lstlisting}
\vspace{-3pt}
\begin{lstlisting}[style=mystyle2, backgroundcolor=\color{yellow!30}]
route-map ISP_OUT permit 10
 match community D2
 match ip address prefix-list D3
 set metric 55
\end{lstlisting}
\vspace{-12pt}
\begin{lstlisting}[style=mystyle2]
route-map ISP_OUT deny 20
 match as-path D0
route-map ISP_OUT deny 30
 match ip address prefix-list D1
route-map  ISP_OUT permit 40
 match local-preference 300
\end{lstlisting}
\vspace{-4pt}
{\color{black}\hrule}
\caption{}
\end{subfigure}\hspace{\fill} 
\begin{subfigure}[t]{0.45\textwidth}
{\color{black}\hrule}
\vspace{-4pt}
    \begin{lstlisting}[style=mystyle2]
ip community-list expanded D2 permit _300:3_
ip prefix-list D3 permit 100.0.0.0/16 le 23

route-map ISP_OUT deny 10
 match as-path D0
route-map ISP_OUT deny 20
 match ip address prefix-list D1
route-map  ISP_OUT permit 30
 match local-preference 300
\end{lstlisting}
\vspace{-12pt}
\begin{lstlisting}[style=mystyle2, backgroundcolor=\color{yellow!30}]
route-map ISP_OUT permit 40
 match community D2
 match ip address prefix-list D3
 set metric 55
\end{lstlisting}
\vspace{-4pt}
{\color{black}\hrule}
\caption{}
\end{subfigure}

\bigskip 
\begin{subfigure}[t]{0.45\textwidth}
{\color{black}\hrule}
\vspace{-4pt}
    \begin{lstlisting}[style=mystyle2]
ip community-list expanded D2 permit _300:3_
ip prefix-list D3 permit 100.0.0.0/16 le 23

route-map ISP_OUT deny 10
 match as-path D0
\end{lstlisting}
\vspace{-12pt}
\begin{lstlisting}[style=mystyle2, backgroundcolor=\color{yellow!30}]
route-map ISP_OUT permit 20
 match community D2
 match ip address prefix-list D3
 set metric 55
\end{lstlisting}
\vspace{-12pt}
\begin{lstlisting}[style=mystyle2]
route-map ISP_OUT deny 30
 match ip address prefix-list D1
route-map  ISP_OUT permit 40
 match local-preference 300
\end{lstlisting}
\vspace{-4pt}
{\color{black}\hrule}
\caption{}
\end{subfigure}\hspace{\fill}
\begin{subfigure}[t]{0.45\textwidth}
{\color{black}\hrule}
\vspace{-4pt}
    \begin{lstlisting}[style=mystyle2]
ip community-list expanded D2 permit _300:3_
ip prefix-list D3 permit 100.0.0.0/16 le 23

route-map ISP_OUT deny 10
 match as-path D0
route-map ISP_OUT deny 20
 match ip address prefix-list D1
\end{lstlisting}
\vspace{-12pt}
\begin{lstlisting}[style=mystyle2, backgroundcolor=\color{yellow!30}]
route-map ISP_OUT permit 30
 match community D2
 match ip address prefix-list D3
 set metric 55
\end{lstlisting}
\vspace{-12pt}
\begin{lstlisting}[style=mystyle2]
route-map  ISP_OUT permit 40
 match local-preference 300
\end{lstlisting}
\vspace{-4pt}
{\color{black}\hrule}
\caption{}
\end{subfigure}

\caption{Possible options for inserting the LLM synthesized stanza into the existing route-map \texttt{ISP\_OUT}. Config snippets highlighted in yellow indicate the location of the newly inserted stanza. AS path list \texttt{D0} and prefix list \texttt{D1} have been omitted from the snippets for brevity. Data structure names are automatically changed by the tool during insertion. 
}
\label{options}
\end{figure*}

\subsection{Disambiguation}
\label{disambiguation}
After verifying syntactic and semantic correctness for the LLM-generated stanza in isolation, the next step is to insert it into the correct location in the original route-map (\mycirc{6} in Figure \ref{workflow}). Figure \ref{options} shows 4 possible insertion scenarios. To help the user determine intent,
we use a tool called a disambiguator that compares different scenarios (using the \texttt{compareRoutePolicies} analysis in Batfish) and helps the user to clarify their preferences through differential examples. Currently, our disambiguator prototype only supports stanza insertions at the top or bottom of the initial route-map (a and b in Figure \ref{options}). For example, it could find the following input route (among others) that leads to differential behavior with respect to the route-maps of Figures \ref{options}(a) and \ref{options}(b):

\vspace{10pt}
{\color{black}\hrule}
\vspace{-4pt}
\begin{lstlisting}[basicstyle=\ttfamily\bfseries]
Network: 100.0.0.0/16
AS Path: [{
        "asns": [32],
        "confederation": false
    }]
Communities: ["300:3"]
Local Preference: 100
Metric: 0
Next Hop IP: 0.0.0.1
Tag: 0
Weight: 0    
\end{lstlisting}
\vspace{-4pt}
{\color{black}\hrule}
\vspace{10pt}

\noindent There are two possible disambiguations.

\noindent\textbf{OPTION 1:}
\vspace{5pt}
{\color{black}\hrule}
\vspace{-4pt}
\begin{lstlisting}[basicstyle=\ttfamily\bfseries]
ACTION: permit
Network: 100.0.0.0/16
AS Path: [{"asns": [32],
        "confederation": false}]
Communities: ["300:3"]
Local Preference: 100
Metric: 55
Next Hop IP: 0.0.0.1
Tag: 0
Weight: 0    
\end{lstlisting}
\vspace{-4pt}
{\color{black}\hrule}
\vspace{15pt}

\noindent\textbf{OPTION 2:}
\vspace{5pt}
{\color{black}\hrule}
\vspace{-4pt}
\begin{lstlisting}[basicstyle=\ttfamily\bfseries]
ACTION: deny
\end{lstlisting}
\vspace{-4pt}
{\color{black}\hrule}
\vspace{10pt}

Note that there are many differences detected by our tool even for this one comparison. But we only show one example here. The user is then asked to select which behavior they want. For the current example, we selected the first option and thus we get the final route-map as shown in Figure \ref{options}(a).

\section{How common are overlaps?}
\label{motivation}

Ambiguity is a real problem only if route maps and ACLs used in practice have considerable overlap.
We developed a Batfish extension to analyze the frequency and scope of overlaps in route-maps within a university campus network and the Wide Area Network (WAN) of a major cloud provider.
Two rules in an ACL are said to have a conflicting overlap if they perform different actions on a packet containing a header that is successfully matched by both. For route-maps, we define two stanzas to have an overlap if there is at least one route advertisement that successfully matches both. We ignore actions for route maps because a route-map stanza may be linked to other route-maps using  \texttt{goto}, \texttt{continue} and  \texttt{call} statements. Thus the route map overlap calculation is an upper bound.

  

\subsection{Overlaps in a Cloud Network}
\label{cloud_overlaps}
We examined 237 non-identical ACLs, some of which may be created from the same template, and determined that 69 had at least one overlap; of these, 48 were found to have an overlap count of more than 20. In one case, an ACL that processes nodes entering the network from an outside network contained dozens of rules permitting and denying combinations of source prefixes, destination prefixes, and protocols. This results in over 100 pairs of overlapping rules. In this case, the ordering of the rules can have a large effect on the behavior of the ACL.

Turning to route maps, the ones applied to routes from or to external neighbors perform fairly complex logic; thus several contain overlapping stanzas. We examined 800 policies, and found 140 contain overlaps. Of these, 3 were found to have more than 20 overlaps each. However, there is an additional source of complexity. In the campus network, each BGP neighbor typically used one route map for importing and one for exporting. In routers we examined in the cloud, it was more common to use a sequence of multiple route maps. Hence, there can be overlaps not just between different stanzas within a single route map, but also between different route maps applied to the same neighbor.

In summary, overlaps are very common making manual incremental changes perilous whether done by an LLM based system or a human operator.  A small error in intent 
can break existing policies and cause major network downtime~\cite{youtube}. 
\subsection{Overlaps in a Campus Network}
\label{campus_overlaps}
In the campus network consisting of 1421 device configurations, we analyzed 169 route-maps. We found 2 route-maps with overlapping stanzas. One route-map \texttt{border\_blackhole} had 3 overlapping stanza pairs, of which 2 were conflicting. 

Access-control lists were more widely used in the campus network with 11,088 ACLs. Of these, 37.7\% had conflicting rule overlaps. Further, 27\% of such ACLs had more than 20 conflicts. This analysis included pairs where one rule's match condition is a proper subset of the other (e.g. \texttt{permit tcp host 1.1.1.1 host 2.2.2.} and \texttt{deny ip any any}). If we ignore such cases, then the percentage of ACLs with non-trivial overlaps comes out to be approximately 18.6\%. Of these, 16.3\% show an overlap count of greater than 20.
\section{Disambiguation Algorithm} 
\label{algo}
In this section we formalize the disambiguation problem and our approach.
Let $Input$ denote the set of all possible input routes or packets to a route map or ACL. We model each of these components abstractly as a sequence of {\em rules}.  Let $Rule$ be the set of all possible rules. A rule $S$ matches against an input $r$ based on specified conditions and performs some action on it. If there is a successful match, then the function $\matches(r,S)$ returns true, otherwise false. 

A route map or ACL is then modeled as a list of rules $\overline{S} = [S_{1}, S_{2}, ..., S_{n}]$, and its semantics is defined by the function $M: Input \rightarrow Rule$ defined as follows:

$$\forall r \in Input, \ M(r) = \mathrm{argmin}_{S \in \overline{S}} \mid \matches(r, S)$$

\vspace{10pt}

The function $M$ formalizes which rule each route is handled by --- the leftmost rule that matches. Note that route maps and ACLs have an implicit deny statement in the end, which we can model by adding an explicit rule at the end of $\overline{S}$ for this purpose.

We model the disambiguation problem as follows. The user would like to insert a new rule, $\mathcal{S*}$, into $\overline{S}$, and the intent is for the resulting list of rules to satisfy a new semantic function $M': Input \rightarrow Rule$. Of course, $M'$ cannot be arbitrary --- it must have a strong relationship to $M$ in order for it to be possible to be constructed solely by inserting a single new rule. We formalize the required relationship between $M$ and $M'$ in the following three conditions that any $M'$ must satisfy:

\begin{itemize}
    \item $\forall r \in Input, \ M'(r) = M(r) \ \vee M'(r) = S*$
    \item $\forall r \in Input, \ M'(r) = S* \Rightarrow \matches(r, S*)$
    \item $\forall r, r' \in Input, \ \matches(r,S*) \ \wedge \ \matches(r',S*) \ \wedge \ M'(r) = M(r) \ \wedge \ M'(r') = S* \Rightarrow M(r) \le M(r')$
\end{itemize}


\begin{figure}
\begin{center}
\begin{tabular}{||c | c | c | c||} 
 \hline
 Router & \#Route-maps & \#LLM calls & \#Disambiguation \\ [0.5ex] 
 \hline\hline
 M & 4 & 9 & 5 \\ 
 \hline
 R1 & 5 & 12 & 6 \\
 \hline
 R2 & 5 & 12 & 6 \\ 
 \hline
\end{tabular}
\end{center}
\caption{Statistics for generating and disambiguating the route-maps for Figure~\ref{fig:topology} incrementally.}
\label{tab:stats}
\end{figure}

The first condition formalizes the incremental nature of the update: every route is either handled as it was before or is handled by the new rule $S*$. The second condition ensures that any route handled by the new rule is in fact matched by that rule. The final condition is perhaps the most interesting: it ensures that there is a single location where $S*$ can be inserted into $\overline{S}$ in order to implement $M'$. Specifically, if there are two inputs $r$ and $r'$ that match the new stanza $S*$ but only $r'$ should be handled by the new stanza, then the original stanza handling $r$ must come before the original stanza handling $r'$, so that we can place $S*$ somewhere in between.

Given these conditions on $M'$, we can use binary search to solve the disambiguation problem and determine where to insert $\mathcal{S*}$ into $\overline{S}$. First, we collect all rules $\{\mathcal{S}\}$ (maintaining their relative order) in $\overline{S}$ for which $\exists r \in Input$ such that $\matches(r, S) \ \wedge \matches(r, S*)$. Then, we obtain the middle rule from this new subset, and ask the user to clarify the desired behavior by showing them a differential example. Based on the user's choice, one half of the subset is discarded and the search continues in the other half. Hence, users are queried a logarithmic number of times to fully disambiguate insertion. 


\section{Evaluation}
\label{evaluation}
\begin{figure}
    \centering
    \includegraphics[width=0.35\textwidth]{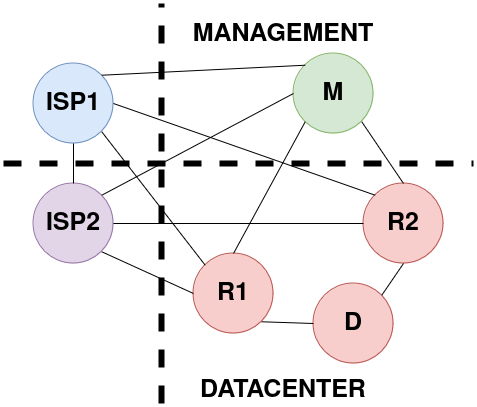}
    \caption{Network topology used for evaluation}
    \label{fig:topology}
\end{figure}
To validate \tool{}'s efficacy in implementing all the router configurations within a network incrementally from scratch, we created a synthetic topology as shown in Figure \ref{fig:topology}, inspired by Lightyear \cite{lightyear} and implemented the following global policies on it:
\begin{itemize}
    \item Reused prefixes within datacenter should not be visible to management
    \item Reused prefixes within management should not be visible to datacenter
    \item The special prefix 10.1.0.0/16 (which is a service within the datacenter) should be visible to M
    \item M should prefer the path through R1 to reach 10.1.0.0/16.
    \item No bogon prefixes should be advertised.
    \item ISP1 and ISP2 should not be reachable to each other through our network.
\end{itemize}
These policies are similar to those used in wide-area networks \cite{lightyear}. Following Lightyear~\cite{lightyear}, we decomposed these global policies into local policies for each router in Figure \ref{fig:topology}, and incrementally synthesized the configurations for $R1$, $R2$ and $M$. Figure \ref{tab:stats} shows the statistics for the number of route-maps per router, number of calls made to the LLM and the number of times the user had to clarify their choice during updates. Some route-maps were reused because similar policies were applied on interfaces, reducing the number of LLM calls.

We found that using the prompting technique of Section~\ref{methodology}, GPT-4 was able to synthesize the correct stanza every time in a single pass and no errors were detected.  While this is promising,
this is a simple topology with only a few stanzas in each route-map. Much more experience is required with real operators.


\section{Related Work}
\label{related_work}
In existing work, synthesis is done in two ways: either by specifying intents in domain-specific languages which are subsequently compiled to concrete device configurations \cite{propane, propane-at, network-wide, aura, snowcap, aed}, or by representing intents as templates which are filled in by a program to satisfy network policies~\cite{netcomplete, robotron}. Both require major human involvement and expertise. The first approach requires the user to be conversant with the DSL. 

Decomposition of global intents to local sub-policies by a black-box compiler makes it difficult for operators to interpret and use across multiple devices. 
Some DSL approaches \cite{propane, network-wide} do not support incremental changes, requiring the complete configurations to be regenerated  even for a small network change. Other approaches address this problem~\cite{snowcap, propane-at, aed}. For template-based synthesis, the network operator must be able to map arbitrary intents to configuration templates, which can be challenging.

To reduce operator effort, generative AI has been proposed for network configuration synthesis \cite{netbuddy, netconfeval, hotnets, preconfig, cegs, genet, intent-based}. 
LLMs can directly translate ambiguous English intents to usable configuration snippets \cite{netbuddy, preconfig, hotnets}. However, the resulting config can have errors. Prior work addresses errors either by including a verifier in the loop \cite{hotnets, alphaverus}, or by pretraining on a networking dataset compiled using a combination of manual and heuristic (even using LLMs) data synthesis, curation and cleaning approaches~\cite{preconfig}. Both these methodologies have been used in the broader domain of software engineering ~\cite{clover,datasynthesis}.
For template-based synthesis, researchers have used graph neural networks~\cite{configreco} to recommend configuration templates based on the input intent, which os then filled manually. 
Automated template generation has also been explored using large language models in \cite{graphsynth}. CEGS \cite{cegs} further automates the template generation process by first retrieving appropriate templates from device documentations and then mapping each template to specific devices in the network and then synthesizing them using NetComplete \cite{netcomplete} which, however, does not support incremental changes. If a user has multiple intents, all of them must be specified at once to output the final device configurations.  
In summary, while a great deal of related work is complementary, it does not address the ambiguity problem in incremental synthesis.

\section{Conclusion and Future Work}
\label{conclusion}

While much research has focused on dealing with hallucinations in LLM produced code, our paper deals with incorrect code produced by ambiguity in user intent.
An LLM generates a formal snippet based on Natural Language instructions, but the snippet has to be inserted into an existing set of snippets where order matters.  The LLM cannot infer the order because there are several equally valid places the new snippet can be placed in.  

\remove{
One may argue that the user could find a way to communicate the intended order to the LLM. Unfortunately, when multiple snippets are added over a very long period (as in a typical router configuration), the user may not remember past overlapping rules: worse, they may have been added by others. Theoretically, this memory of past rules can be outsourced to the LLM. However, this requires a very large context window which is limited even in today's most powerful LLMs.  Moreover, this is really asking the LLM to behave like a database, which is not its strength. Thus it makes sense to outsource the database of past rules, and the finding of overlaps with past rules, to a separate module we call a Disambiguator. The Disambiguator also elicits the right order from the user by giving it examples of differential behavior for various orders.
}

A natural question is whether the LLM itself could play the role of the disambiguator. While this would be possible to explore, we believe that symbolic reasoning tools are a better fit, since at that point in the process we have structured inputs with well-defined semantics (candidate configuration updates) and a precise goal (identifying behavioral differences). Our approach still leverages the LLM for what it does better than any other technology -- turning natural language specifications into configuration stanzas.  

Our paper is only a toy demonstration, and
there is much work to be done to make it a practical tool
for network configuration updates. First, the tool needs support for inserting entries into other data structures that can have conflicts like prefix lists, access-lists, community-lists and AS-path lists. Second, the disambiguator currently only handles two insertion locations. Third, we have only used one LLM augmentation technique (few-shot examples).  Can using chain-of-thought, retrieval-augemented generation, graph RAG or even agentic AI do better?


Finally, we note that the problem of intent disambiguation is very general, so our solution is potentially applicable to other settings. For example, disambiguation is required for code generation tasks beyond network configurations, and it
can be useful for network configuration updates even if they are performed manually rather than via an LLM.

\section {Acknowledgement}

The work of Rajdeep Mondal, Todd Millstein and George Varghese at the University of California Los Angeles (UCLA) was partly supported by the National Science Foundation (NSF) under award CNS-2402958.

\bibliographystyle{ACM-Reference-Format} 
\bibliography{hotnets25-template}


\begin{thebibliography}{32}


\ifx \showCODEN    \undefined \def \showCODEN     #1{\unskip}     \fi
\ifx \showDOI      \undefined \def \showDOI       #1{#1}\fi
\ifx \showISBNx    \undefined \def \showISBNx     #1{\unskip}     \fi
\ifx \showISBNxiii \undefined \def \showISBNxiii  #1{\unskip}     \fi
\ifx \showISSN     \undefined \def \showISSN      #1{\unskip}     \fi
\ifx \showLCCN     \undefined \def \showLCCN      #1{\unskip}     \fi
\ifx \shownote     \undefined \def \shownote      #1{#1}          \fi
\ifx \showarticletitle \undefined \def \showarticletitle #1{#1}   \fi
\ifx \showURL      \undefined \def \showURL       {\relax}        \fi
\providecommand\bibfield[2]{#2}
\providecommand\bibinfo[2]{#2}
\providecommand\natexlab[1]{#1}
\providecommand\showeprint[2][]{arXiv:#2}

\bibitem[Abhashkumar et~al\mbox{.}(2020)]%
        {aed}
\bibfield{author}{\bibinfo{person}{Anubhavnidhi Abhashkumar}, \bibinfo{person}{Aaron Gember-Jacobson}, {and} \bibinfo{person}{Aditya Akella}.} \bibinfo{year}{2020}\natexlab{}.
\newblock \showarticletitle{AED: incrementally synthesizing policy-compliant and manageable configurations}. In \bibinfo{booktitle}{\emph{Proceedings of the 16th International Conference on Emerging Networking EXperiments and Technologies}} (Barcelona, Spain) \emph{(\bibinfo{series}{CoNEXT '20})}. \bibinfo{publisher}{Association for Computing Machinery}, \bibinfo{address}{New York, NY, USA}, \bibinfo{pages}{482–495}.
\newblock
\showISBNx{9781450379489}
\urldef\tempurl%
\url{https://doi.org/10.1145/3386367.3431304}
\showDOI{\tempurl}


\bibitem[Asai et~al\mbox{.}(2023)]%
        {selfrag}
\bibfield{author}{\bibinfo{person}{Akari Asai}, \bibinfo{person}{Zeqiu Wu}, \bibinfo{person}{Yizhong Wang}, \bibinfo{person}{Avirup Sil}, {and} \bibinfo{person}{Hannaneh Hajishirzi}.} \bibinfo{year}{2023}\natexlab{}.
\newblock \bibinfo{title}{Self-RAG: Learning to Retrieve, Generate, and Critique through Self-Reflection}.
\newblock
\newblock
\showeprint[arxiv]{2310.11511}~[cs.CL]
\urldef\tempurl%
\url{https://arxiv.org/abs/2310.11511}
\showURL{%
\tempurl}


\bibitem[Beckett et~al\mbox{.}(2016)]%
        {propane}
\bibfield{author}{\bibinfo{person}{Ryan Beckett}, \bibinfo{person}{Ratul Mahajan}, \bibinfo{person}{Todd Millstein}, \bibinfo{person}{Jitendra Padhye}, {and} \bibinfo{person}{David Walker}.} \bibinfo{year}{2016}\natexlab{}.
\newblock \showarticletitle{Don't Mind the Gap: Bridging Network-wide Objectives and Device-level Configurations}. In \bibinfo{booktitle}{\emph{Proceedings of the 2016 ACM SIGCOMM Conference}} (Florianopolis, Brazil) \emph{(\bibinfo{series}{SIGCOMM '16})}. \bibinfo{publisher}{Association for Computing Machinery}, \bibinfo{address}{New York, NY, USA}, \bibinfo{pages}{328–341}.
\newblock
\showISBNx{9781450341936}
\urldef\tempurl%
\url{https://doi.org/10.1145/2934872.2934909}
\showDOI{\tempurl}


\bibitem[Beckett et~al\mbox{.}(2017)]%
        {propane-at}
\bibfield{author}{\bibinfo{person}{Ryan Beckett}, \bibinfo{person}{Ratul Mahajan}, \bibinfo{person}{Todd Millstein}, \bibinfo{person}{Jitendra Padhye}, {and} \bibinfo{person}{David Walker}.} \bibinfo{year}{2017}\natexlab{}.
\newblock \showarticletitle{Network configuration synthesis with abstract topologies}.
\newblock \bibinfo{journal}{\emph{SIGPLAN Not.}} \bibinfo{volume}{52}, \bibinfo{number}{6} (\bibinfo{date}{June} \bibinfo{year}{2017}), \bibinfo{pages}{437–451}.
\newblock
\showISSN{0362-1340}
\urldef\tempurl%
\url{https://doi.org/10.1145/3140587.3062367}
\showDOI{\tempurl}


\bibitem[Edge et~al\mbox{.}(2025)]%
        {graphrag}
\bibfield{author}{\bibinfo{person}{Darren Edge}, \bibinfo{person}{Ha Trinh}, \bibinfo{person}{Newman Cheng}, \bibinfo{person}{Joshua Bradley}, \bibinfo{person}{Alex Chao}, \bibinfo{person}{Apurva Mody}, \bibinfo{person}{Steven Truitt}, \bibinfo{person}{Dasha Metropolitansky}, \bibinfo{person}{Robert~Osazuwa Ness}, {and} \bibinfo{person}{Jonathan Larson}.} \bibinfo{year}{2025}\natexlab{}.
\newblock \bibinfo{title}{From Local to Global: A Graph RAG Approach to Query-Focused Summarization}.
\newblock
\newblock
\showeprint[arxiv]{2404.16130}~[cs.CL]
\urldef\tempurl%
\url{https://arxiv.org/abs/2404.16130}
\showURL{%
\tempurl}


\bibitem[El{-}Hassany et~al\mbox{.}(2017)]%
        {network-wide}
\bibfield{author}{\bibinfo{person}{Ahmed El{-}Hassany}, \bibinfo{person}{Petar Tsankov}, \bibinfo{person}{Laurent Vanbever}, {and} \bibinfo{person}{Martin~T. Vechev}.} \bibinfo{year}{2017}\natexlab{}.
\newblock \showarticletitle{Network-Wide Configuration Synthesis}. In \bibinfo{booktitle}{\emph{Computer Aided Verification - 29th International Conference, {CAV} 2017, Heidelberg, Germany, July 24-28, 2017, Proceedings, Part {II}}} \emph{(\bibinfo{series}{Lecture Notes in Computer Science}, Vol.~\bibinfo{volume}{10427})}, \bibfield{editor}{\bibinfo{person}{Rupak Majumdar} {and} \bibinfo{person}{Viktor Kuncak}} (Eds.). \bibinfo{publisher}{Springer}, \bibinfo{pages}{261--281}.
\newblock
\urldef\tempurl%
\url{https://doi.org/10.1007/978-3-319-63390-9\_14}
\showDOI{\tempurl}


\bibitem[Fuad et~al\mbox{.}(2024)]%
        {intent-based}
\bibfield{author}{\bibinfo{person}{Ahlam Fuad}, \bibinfo{person}{Azza~H. Ahmed}, \bibinfo{person}{Michael~A. Riegler}, {and} \bibinfo{person}{Tarik Čičić}.} \bibinfo{year}{2024}\natexlab{}.
\newblock \showarticletitle{An Intent-based Networks Framework based on Large Language Models}. In \bibinfo{booktitle}{\emph{2024 IEEE 10th International Conference on Network Softwarization (NetSoft)}}. \bibinfo{pages}{7--12}.
\newblock
\urldef\tempurl%
\url{https://doi.org/10.1109/NetSoft60951.2024.10588879}
\showDOI{\tempurl}


\bibitem[Gao et~al\mbox{.}(2024)]%
        {modularrag}
\bibfield{author}{\bibinfo{person}{Yunfan Gao}, \bibinfo{person}{Yun Xiong}, \bibinfo{person}{Meng Wang}, {and} \bibinfo{person}{Haofen Wang}.} \bibinfo{year}{2024}\natexlab{}.
\newblock \bibinfo{title}{Modular RAG: Transforming RAG Systems into LEGO-like Reconfigurable Frameworks}.
\newblock
\newblock
\showeprint[arxiv]{2407.21059}~[cs.CL]
\urldef\tempurl%
\url{https://arxiv.org/abs/2407.21059}
\showURL{%
\tempurl}


\bibitem[Guo et~al\mbox{.}(2024a)]%
        {netcomplete}
\bibfield{author}{\bibinfo{person}{Zhenbei Guo}, \bibinfo{person}{Fuliang Li}, \bibinfo{person}{Jiaxing Shen}, \bibinfo{person}{Tangzheng Xie}, \bibinfo{person}{Shan Jiang}, {and} \bibinfo{person}{Xingwei Wang}.} \bibinfo{year}{2024}\natexlab{a}.
\newblock \showarticletitle{ConfigReco: Network Configuration Recommendation With Graph Neural Networks}.
\newblock \bibinfo{journal}{\emph{IEEE Network}} \bibinfo{volume}{38}, \bibinfo{number}{1} (\bibinfo{year}{2024}), \bibinfo{pages}{7--14}.
\newblock
\urldef\tempurl%
\url{https://doi.org/10.1109/MNET.2023.3336239}
\showDOI{\tempurl}


\bibitem[Guo et~al\mbox{.}(2024b)]%
        {configreco}
\bibfield{author}{\bibinfo{person}{Zhenbei Guo}, \bibinfo{person}{Fuliang Li}, \bibinfo{person}{Jiaxing Shen}, \bibinfo{person}{Tangzheng Xie}, \bibinfo{person}{Shan Jiang}, {and} \bibinfo{person}{Xingwei Wang}.} \bibinfo{year}{2024}\natexlab{b}.
\newblock \showarticletitle{ConfigReco: Network Configuration Recommendation With Graph Neural Networks}.
\newblock \bibinfo{journal}{\emph{IEEE Network}} \bibinfo{volume}{38}, \bibinfo{number}{1} (\bibinfo{year}{2024}), \bibinfo{pages}{7--14}.
\newblock
\urldef\tempurl%
\url{https://doi.org/10.1109/MNET.2023.3336239}
\showDOI{\tempurl}


\bibitem[Hu et~al\mbox{.}(2025)]%
        {AgenticAI}
\bibfield{author}{\bibinfo{person}{Yaojie Hu}, \bibinfo{person}{Qiang Zhou}, \bibinfo{person}{Qihong Chen}, \bibinfo{person}{Xiaopeng Li}, \bibinfo{person}{Linbo Liu}, \bibinfo{person}{Dejiao Zhang}, \bibinfo{person}{Amit Kachroo}, \bibinfo{person}{Talha Oz}, {and} \bibinfo{person}{Omer Tripp}.} \bibinfo{year}{2025}\natexlab{}.
\newblock \bibinfo{title}{QualityFlow: An Agentic Workflow for Program Synthesis Controlled by LLM Quality Checks}.
\newblock
\newblock
\urldef\tempurl%
\url{https://doi.org/10.48550/arXiv.2501.17167}
\showDOI{\tempurl}


\bibitem[Ifland et~al\mbox{.}(2024)]%
        {genet}
\bibfield{author}{\bibinfo{person}{Beni Ifland}, \bibinfo{person}{Elad Duani}, \bibinfo{person}{Rubin Krief}, \bibinfo{person}{Miro Ohana}, \bibinfo{person}{Aviram Zilberman}, \bibinfo{person}{Andres Murillo}, \bibinfo{person}{Ofir Manor}, \bibinfo{person}{Ortal Lavi}, \bibinfo{person}{Hikichi Kenji}, \bibinfo{person}{Asaf Shabtai}, \bibinfo{person}{Yuval Elovici}, {and} \bibinfo{person}{Rami Puzis}.} \bibinfo{year}{2024}\natexlab{}.
\newblock \bibinfo{title}{GeNet: A Multimodal LLM-Based Co-Pilot for Network Topology and Configuration}.
\newblock
\newblock
\showeprint[arxiv]{2407.08249}~[cs.NI]
\urldef\tempurl%
\url{https://arxiv.org/abs/2407.08249}
\showURL{%
\tempurl}


\bibitem[Lewis et~al\mbox{.}(2020)]%
        {rag}
\bibfield{author}{\bibinfo{person}{Patrick S.~H. Lewis}, \bibinfo{person}{Ethan Perez}, \bibinfo{person}{Aleksandra Piktus}, \bibinfo{person}{Fabio Petroni}, \bibinfo{person}{Vladimir Karpukhin}, \bibinfo{person}{Naman Goyal}, \bibinfo{person}{Heinrich K{\"{u}}ttler}, \bibinfo{person}{Mike Lewis}, \bibinfo{person}{Wen{-}tau Yih}, \bibinfo{person}{Tim Rockt{\"{a}}schel}, \bibinfo{person}{Sebastian Riedel}, {and} \bibinfo{person}{Douwe Kiela}.} \bibinfo{year}{2020}\natexlab{}.
\newblock \showarticletitle{Retrieval-Augmented Generation for Knowledge-Intensive {NLP} Tasks}.
\newblock \bibinfo{journal}{\emph{CoRR}}  \bibinfo{volume}{abs/2005.11401} (\bibinfo{year}{2020}).
\newblock
\showeprint[arXiv]{2005.11401}
\urldef\tempurl%
\url{https://arxiv.org/abs/2005.11401}
\showURL{%
\tempurl}


\bibitem[Li et~al\mbox{.}(2024)]%
        {preconfig}
\bibfield{author}{\bibinfo{person}{Fuliang Li}, \bibinfo{person}{Haozhi Lang}, \bibinfo{person}{Jiajie Zhang}, \bibinfo{person}{Jiaxing Shen}, {and} \bibinfo{person}{Xingwei Wang}.} \bibinfo{year}{2024}\natexlab{}.
\newblock \bibinfo{title}{PreConfig: A Pretrained Model for Automating Network Configuration}.
\newblock
\newblock
\showeprint[arxiv]{2403.09369}~[cs.NI]
\urldef\tempurl%
\url{https://arxiv.org/abs/2403.09369}
\showURL{%
\tempurl}


\bibitem[Liu et~al\mbox{.}(2023)]%
        {berkeley}
\bibfield{author}{\bibinfo{person}{Alisa Liu}, \bibinfo{person}{Zhaofeng Wu}, \bibinfo{person}{Julian Michael}, \bibinfo{person}{Alane Suhr}, \bibinfo{person}{Peter West}, \bibinfo{person}{Alexander Koller}, \bibinfo{person}{Swabha Swayamdipta}, \bibinfo{person}{Noah~A. Smith}, {and} \bibinfo{person}{Yejin Choi}.} \bibinfo{year}{2023}\natexlab{}.
\newblock \showarticletitle{We{'}re Afraid Language Models Aren{'}t Modeling Ambiguity}. In \bibinfo{booktitle}{\emph{Proceedings of the 2023 Conference on Empirical Methods in Natural Language Processing}}. \bibinfo{publisher}{Association for Computational Linguistics}, \bibinfo{address}{Singapore}, \bibinfo{pages}{790--807}.
\newblock
\urldef\tempurl%
\url{https://doi.org/10.18653/v1/2023.emnlp-main.51}
\showDOI{\tempurl}


\bibitem[Liu et~al\mbox{.}(2025)]%
        {cegs}
\bibfield{author}{\bibinfo{person}{Jianmin Liu}, \bibinfo{person}{Li Chen}, \bibinfo{person}{Dan Li}, {and} \bibinfo{person}{Yukai Miao}.} \bibinfo{year}{2025}\natexlab{}.
\newblock \showarticletitle{{CEGS}: Configuration Example Generalizing Synthesizer}. In \bibinfo{booktitle}{\emph{22nd USENIX Symposium on Networked Systems Design and Implementation (NSDI 25)}}. \bibinfo{publisher}{USENIX Association}, \bibinfo{address}{Philadelphia, PA}, \bibinfo{pages}{1327--1347}.
\newblock
\showISBNx{978-1-939133-46-5}
\urldef\tempurl%
\url{https://www.usenix.org/conference/nsdi25/presentation/liu-jianmin}
\showURL{%
\tempurl}


\bibitem[McCullagh({[n.\,d.]})]%
        {youtube}
\bibfield{author}{\bibinfo{person}{Declan McCullagh}.} \bibinfo{year}{[n.\,d.]}\natexlab{}.
\newblock \bibinfo{title}{How Pakistan knocked YouTube offline (and how to make sure it never happens again)}.
\newblock
\newblock


\bibitem[Mondal et~al\mbox{.}(2023)]%
        {hotnets}
\bibfield{author}{\bibinfo{person}{Rajdeep Mondal}, \bibinfo{person}{Alan Tang}, \bibinfo{person}{Ryan Beckett}, \bibinfo{person}{Todd Millstein}, {and} \bibinfo{person}{George Varghese}.} \bibinfo{year}{2023}\natexlab{}.
\newblock \showarticletitle{What do LLMs need to Synthesize Correct Router Configurations?} \emph{(\bibinfo{series}{HotNets '23})}. \bibinfo{publisher}{Association for Computing Machinery}, \bibinfo{address}{New York, NY, USA}, \bibinfo{pages}{189–195}.
\newblock
\showISBNx{9798400704154}
\urldef\tempurl%
\url{https://doi.org/10.1145/3626111.3628194}
\showDOI{\tempurl}


\bibitem[Pranjal~Aggarwal(2024)]%
        {alphaverus}
\bibfield{author}{\bibinfo{person}{Sean~Welleck Pranjal~Aggarwal, Bryan~Parno}.} \bibinfo{year}{2024}\natexlab{}.
\newblock \bibinfo{title}{AlphaVerus: Bootstrapping Formally Verified Code Generation through Self-Improving Translation and Treefinement}.
\newblock
\newblock
\showeprint[arxiv]{2405.19616}~[cs.AI]
\urldef\tempurl%
\url{https://arxiv.org/abs/2412.06176}
\showURL{%
\tempurl}


\bibitem[Ramanathan et~al\mbox{.}(2023)]%
        {aura}
\bibfield{author}{\bibinfo{person}{Sivaramakrishnan Ramanathan}, \bibinfo{person}{Ying Zhang}, \bibinfo{person}{Mohab Gawish}, \bibinfo{person}{Yogesh Mundada}, \bibinfo{person}{Zhaodong Wang}, \bibinfo{person}{Sangki Yun}, \bibinfo{person}{Eric Lippert}, \bibinfo{person}{Walid Taha}, \bibinfo{person}{Minlan Yu}, {and} \bibinfo{person}{Jelena Mirkovic}.} \bibinfo{year}{2023}\natexlab{}.
\newblock \showarticletitle{Practical Intent-driven Routing Configuration Synthesis}. In \bibinfo{booktitle}{\emph{20th USENIX Symposium on Networked Systems Design and Implementation (NSDI 23)}}. \bibinfo{publisher}{USENIX Association}, \bibinfo{address}{Boston, MA}, \bibinfo{pages}{629--644}.
\newblock
\showISBNx{978-1-939133-33-5}
\urldef\tempurl%
\url{https://www.usenix.org/conference/nsdi23/presentation/ramanathan}
\showURL{%
\tempurl}


\bibitem[Schneider et~al\mbox{.}(2021)]%
        {snowcap}
\bibfield{author}{\bibinfo{person}{Tibor Schneider}, \bibinfo{person}{R\"{u}diger Birkner}, {and} \bibinfo{person}{Laurent Vanbever}.} \bibinfo{year}{2021}\natexlab{}.
\newblock \showarticletitle{Snowcap: synthesizing network-wide configuration updates}. In \bibinfo{booktitle}{\emph{Proceedings of the 2021 ACM SIGCOMM 2021 Conference}} (Virtual Event, USA) \emph{(\bibinfo{series}{SIGCOMM '21})}. \bibinfo{publisher}{Association for Computing Machinery}, \bibinfo{address}{New York, NY, USA}, \bibinfo{pages}{33–49}.
\newblock
\showISBNx{9781450383837}
\urldef\tempurl%
\url{https://doi.org/10.1145/3452296.3472915}
\showDOI{\tempurl}


\bibitem[Sedova et~al\mbox{.}(2024)]%
        {selfconsistency}
\bibfield{author}{\bibinfo{person}{Anastasiia Sedova}, \bibinfo{person}{Robert Litschko}, \bibinfo{person}{Diego Frassinelli}, \bibinfo{person}{Benjamin Roth}, {and} \bibinfo{person}{Barbara Plank}.} \bibinfo{year}{2024}\natexlab{}.
\newblock \bibinfo{title}{To Know or Not To Know? Analyzing Self-Consistency of Large Language Models under Ambiguity}.
\newblock
\newblock
\showeprint[arxiv]{2407.17125}~[cs.CL]
\urldef\tempurl%
\url{https://arxiv.org/abs/2407.17125}
\showURL{%
\tempurl}


\bibitem[Singh et~al\mbox{.}(2025)]%
        {agenticrag}
\bibfield{author}{\bibinfo{person}{Aditi Singh}, \bibinfo{person}{Abul Ehtesham}, \bibinfo{person}{Saket Kumar}, {and} \bibinfo{person}{Tala~Talaei Khoei}.} \bibinfo{year}{2025}\natexlab{}.
\newblock \bibinfo{title}{Agentic Retrieval-Augmented Generation: A Survey on Agentic RAG}.
\newblock
\newblock
\showeprint[arxiv]{2501.09136}~[cs.AI]
\urldef\tempurl%
\url{https://arxiv.org/abs/2501.09136}
\showURL{%
\tempurl}


\bibitem[Sun et~al\mbox{.}(2024)]%
        {clover}
\bibfield{author}{\bibinfo{person}{Chuyue Sun}, \bibinfo{person}{Ying Sheng}, \bibinfo{person}{Oded Padon}, {and} \bibinfo{person}{Clark Barrett}.} \bibinfo{year}{2024}\natexlab{}.
\newblock \showarticletitle{Clover: Closed-Loop Verifiable Code Generation}. In \bibinfo{booktitle}{\emph{Proceedings of the First International Symposium on AI Verification (SAIV '24)}}. \bibinfo{publisher}{Springer-Verlag}, \bibinfo{pages}{134--155}.
\newblock
\urldef\tempurl%
\url{https://doi.org/10.1007/978-3-031-65112-0_7}
\showDOI{\tempurl}
\newblock
\shownote{Montreal, Canada}.


\bibitem[Sung et~al\mbox{.}(2016)]%
        {robotron}
\bibfield{author}{\bibinfo{person}{Yu-Wei~Eric Sung}, \bibinfo{person}{Xiaozheng Tie}, \bibinfo{person}{Starsky~H.Y. Wong}, {and} \bibinfo{person}{Hongyi Zeng}.} \bibinfo{year}{2016}\natexlab{}.
\newblock \showarticletitle{Robotron: Top-down Network Management at Facebook Scale}. In \bibinfo{booktitle}{\emph{Proceedings of the 2016 ACM SIGCOMM Conference}} (Florianopolis, Brazil) \emph{(\bibinfo{series}{SIGCOMM '16})}. \bibinfo{publisher}{Association for Computing Machinery}, \bibinfo{address}{New York, NY, USA}, \bibinfo{pages}{426–439}.
\newblock
\showISBNx{9781450341936}
\urldef\tempurl%
\url{https://doi.org/10.1145/2934872.2934874}
\showDOI{\tempurl}


\bibitem[Tang et~al\mbox{.}(2023)]%
        {lightyear}
\bibfield{author}{\bibinfo{person}{Alan Tang}, \bibinfo{person}{Ryan Beckett}, \bibinfo{person}{Steven Benaloh}, \bibinfo{person}{Karthick Jayaraman}, \bibinfo{person}{Tejas Patil}, \bibinfo{person}{Todd Millstein}, {and} \bibinfo{person}{George Varghese}.} \bibinfo{year}{2023}\natexlab{}.
\newblock \showarticletitle{Lightyear: Using Modularity to Scale BGP Control Plane Verification}. In \bibinfo{booktitle}{\emph{Proceedings of the ACM SIGCOMM 2023 Conference}} \emph{(\bibinfo{series}{ACM SIGCOMM ’23})}. \bibinfo{publisher}{ACM}.
\newblock
\urldef\tempurl%
\url{https://doi.org/10.1145/3603269.3604842}
\showDOI{\tempurl}


\bibitem[Wang et~al\mbox{.}(2024a)]%
        {netconfeval}
\bibfield{author}{\bibinfo{person}{Changjie Wang}, \bibinfo{person}{Mariano Scazzariello}, \bibinfo{person}{Alireza Farshin}, \bibinfo{person}{Simone Ferlin}, \bibinfo{person}{Dejan Kostic}, {and} \bibinfo{person}{Marco Chiesa}.} \bibinfo{year}{2024}\natexlab{a}.
\newblock \showarticletitle{NetConfEval: Can LLMs Facilitate Network Configuration?}
\newblock \bibinfo{journal}{\emph{Proc. ACM Netw.}} \bibinfo{volume}{2}, \bibinfo{number}{CoNEXT2} (\bibinfo{year}{2024}), \bibinfo{pages}{7:1--7:25}.
\newblock
\urldef\tempurl%
\url{http://dblp.uni-trier.de/db/journals/pacmnet/pacmnet2.html#WangSFFKC24}
\showURL{%
\tempurl}


\bibitem[Wang et~al\mbox{.}(2023)]%
        {netbuddy}
\bibfield{author}{\bibinfo{person}{Changjie Wang}, \bibinfo{person}{Mariano Scazzariello}, \bibinfo{person}{Alireza Farshin}, \bibinfo{person}{Dejan Kostic}, {and} \bibinfo{person}{Marco Chiesa}.} \bibinfo{year}{2023}\natexlab{}.
\newblock \bibinfo{title}{Making Network Configuration Human Friendly}.
\newblock
\newblock
\showeprint[arxiv]{2309.06342}~[cs.NI]
\urldef\tempurl%
\url{https://arxiv.org/abs/2309.06342}
\showURL{%
\tempurl}


\bibitem[Wang et~al\mbox{.}(2024b)]%
        {datasynthesis}
\bibfield{author}{\bibinfo{person}{Ke Wang}, \bibinfo{person}{Jiahui Zhu}, \bibinfo{person}{Minjie Ren}, \bibinfo{person}{Zeming Liu}, \bibinfo{person}{Shiwei Li}, \bibinfo{person}{Zongye Zhang}, \bibinfo{person}{Chenkai Zhang}, \bibinfo{person}{Xiaoyu Wu}, \bibinfo{person}{Qiqi Zhan}, \bibinfo{person}{Qingjie Liu}, {and} \bibinfo{person}{Yunhong Wang}.} \bibinfo{year}{2024}\natexlab{b}.
\newblock \bibinfo{title}{A Survey on Data Synthesis and Augmentation for Large Language Models}.
\newblock
\newblock
\showeprint[arxiv]{2410.12896}~[cs.CL]
\urldef\tempurl%
\url{https://arxiv.org/abs/2410.12896}
\showURL{%
\tempurl}


\bibitem[Wang et~al\mbox{.}(2025)]%
        {speculativerag}
\bibfield{author}{\bibinfo{person}{Zilong Wang}, \bibinfo{person}{Zifeng Wang}, \bibinfo{person}{Long Le}, \bibinfo{person}{Steven Zheng}, \bibinfo{person}{Swaroop Mishra}, \bibinfo{person}{Vincent Perot}, \bibinfo{person}{Yuwei Zhang}, \bibinfo{person}{Anush Mattapalli}, \bibinfo{person}{Ankur Taly}, \bibinfo{person}{Jingbo Shang}, \bibinfo{person}{Chen-Yu Lee}, {and} \bibinfo{person}{Tomas Pfister}.} \bibinfo{year}{2025}\natexlab{}.
\newblock \showarticletitle{Speculative {RAG}: Enhancing Retrieval Augmented Generation through Drafting}. In \bibinfo{booktitle}{\emph{The Thirteenth International Conference on Learning Representations}}.
\newblock
\urldef\tempurl%
\url{https://openreview.net/forum?id=xgQfWbV6Ey}
\showURL{%
\tempurl}


\bibitem[Wei et~al\mbox{.}(2023)]%
        {chain_of_thought}
\bibfield{author}{\bibinfo{person}{Jason Wei}, \bibinfo{person}{Xuezhi Wang}, \bibinfo{person}{Dale Schuurmans}, \bibinfo{person}{Maarten Bosma}, \bibinfo{person}{Brian Ichter}, \bibinfo{person}{Fei Xia}, \bibinfo{person}{Ed Chi}, \bibinfo{person}{Quoc Le}, {and} \bibinfo{person}{Denny Zhou}.} \bibinfo{year}{2023}\natexlab{}.
\newblock \bibinfo{title}{Chain-of-Thought Prompting Elicits Reasoning in Large Language Models}.
\newblock
\newblock
\showeprint[arxiv]{2201.11903}~[cs.CL]
\urldef\tempurl%
\url{https://arxiv.org/abs/2201.11903}
\showURL{%
\tempurl}


\bibitem[Zhang et~al\mbox{.}(2025)]%
        {graphsynth}
\bibfield{author}{\bibinfo{person}{Xiaofeng Zhang}, \bibinfo{person}{Xianming Gao}, \bibinfo{person}{Peilin Tao}, {and} \bibinfo{person}{Tao Feng}.} \bibinfo{year}{2025}\natexlab{}.
\newblock \bibinfo{booktitle}{\emph{GraphSynth: Synthesis of Network Configuration Templates Using Large Language Models}}.
\newblock \bibinfo{publisher}{Association for Computing Machinery}, \bibinfo{address}{New York, NY, USA}, \bibinfo{pages}{108–114}.
\newblock
\showISBNx{9798400713453}
\urldef\tempurl%
\url{https://doi.org/10.1145/3728725.3728742}
\showURL{%
\tempurl}


\end{thebibliography}

\end{document}